\documentclass{article}

\usepackage{arxiv}

\usepackage[utf8]{inputenc} 
\usepackage[T1]{fontenc}    
\usepackage{hyperref}       
\usepackage{url}            
\usepackage{booktabs}       
\usepackage{amsfonts}       
\usepackage{nicefrac}       
\usepackage{microtype}      
\usepackage{cleveref}       
\usepackage{lipsum}         
\usepackage{graphicx}
\usepackage{natbib}
\usepackage{doi}
\usepackage{tikz}
\usepackage{multirow}
\usepackage{xcolor}
\usepackage{listings}

\usetikzlibrary{shapes,shadows,positioning,fit,calc}
\tikzset{
  smalltable/.style={
    draw,
    very thin,
    minimum width=1.8cm,
    minimum height=1.2cm,
    inner sep=0,
    grid step/.store in=\stGridStep,
    grid step=3mm,
    grid color/.store in=\stGridColor,
    grid color=black!25,
    path picture={
      \begin{scope}
        \clip (path picture bounding box.south west)
              rectangle (path picture bounding box.north east);
        \draw[\stGridColor, very thin, step=\stGridStep]
          (path picture bounding box.south west)
            grid
          (path picture bounding box.north east);
      \end{scope}
    }
  }
}

\title{Extending GouDa: Generation of Universal Datasets with (and without) Errors for Data Quality Benchmarking}

\date{}

\newif\ifuniqueAffiliation

\ifuniqueAffiliation 
\author{ 
    \href{https://orcid.org/0000-0002-5960-5886}{\includegraphics[scale=0.06]{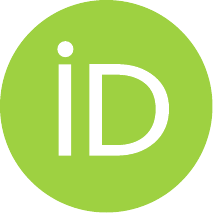}\hspace{1mm}Valerie~Restat} \\
	FernUniversität in Hagen\\
    Hagen, Germany\\
	\href{mailto:valerie.restat@fernuni-hagen.de}{\texttt{valerie.restat@fernuni-hagen.de}}\\
	\AND
    \href{https://orcid.org/0009-0003-7595-532X}{\includegraphics[scale=0.06]{orcid.pdf}\hspace{1mm}Kevin M.~Kramer}\\
	FernUniversität in Hagen\\
    Hagen, Germany\\
	\href{mailto:kevin.kramer@fernuni-hagen.de}{\texttt{kevin.kramer@fernuni-hagen.de}}\\
	\And
	\href{https://orcid.org/0000-0001-6681-2798}{\includegraphics[scale=0.06]{orcid.pdf}\hspace{1mm}André~Conrad} \\
	FernUniversität in Hagen\\
    Hagen, Germany\\
	\href{mailto:andre.conrad@fernuni-hagen.de}{\texttt{andre.conrad@fernuni-hagen.de}}\\
	\And
	\href{https://orcid.org/0000-0003-2771-142X}{\includegraphics[scale=0.06]{orcid.pdf}\hspace{1mm}Uta~Störl} \\
	FernUniversität in Hagen\\
    Hagen, Germany\\
	\href{mailto:uta.stoerl@fernuni-hagen.de}{\texttt{uta.stoerl@fernuni-hagen.de}}\\
}
\else
\usepackage{authblk}

\newbox{\orcid}\sbox{\orcid}{\includegraphics[scale=0.06]{orcid.pdf}} 
\author[1]{%
	\href{https://orcid.org/0000-0002-5960-5886}{\usebox{\orcid}\hspace{1mm}Valerie~Restat\thanks{Email address of the corresponding author: \texttt{valerie.restat@fernuni-hagen.de}}}%
}
\author[1]{%
	\href{https://orcid.org/0000-0001-6681-2798}{\usebox{\orcid}\hspace{1mm}André~Conrad}%
}
\author[1]{%
	\href{https://orcid.org/0009-0003-7595-532X}{\usebox{\orcid}\hspace{1mm}Kevin M.~Kramer}%
}
\author[1]{%
	\href{https://orcid.org/0000-0003-2771-142X}{\usebox{\orcid}\hspace{1mm}Uta~Störl}%
}

\affil[1]{Faculty of Mathematics and Computer Science, FernUniversität in Hagen, Hagen, Germany}

\renewcommand{\headeright}

\hypersetup{
pdftitle={Extending GouDa: Generation of Universal Datasets with (and without) Errors for Data Quality Benchmarking},
pdfsubject={cd.db},
pdfauthor={Valerie~Restat, André Conrad, Kevin M.~Kramer, Uta~Störl},
pdfkeywords={data quality, data generation, evaluation, data sets},
}

\begin{document}
\maketitle
\vspace{-1em}
\begin{abstract}
	Synthetic data is extremely important in areas such as data quality, data cleaning, and machine learning. It enables the analysis of use cases in which real data is insufficient, unavailable, or distorted. However, generating synthetic data also presents challenges: The data must be as realistic as possible, but at the same time cover edge cases. It must be possible to insert controlled errors, and at the same time, an error-free version of the data is usually required. Additionally, it is necessary to consider numerous data formats, such as tabular data, but also NoSQL data models. To this end, we present our data generator GouDa. GouDa precisely meets these requirements -- it is suitable for different data formats, enables the controlled insertion of errors, and generates ground truth. A wide range of different generation functions and the option to add your own lists of possible attribute values allow the generation of realistic data that covers many different use cases.
\end{abstract}

\section{Introduction}
Data is a central concern of many business and scientific problems. But data is often not available in sufficient quality. Poor data quality can lead to wrong decisions and customer frustration, resulting in significant additional costs~\cite{Ehrlinger2025}. Industry studies often estimate that the costs resulting from poor data quality are in the millions~\cite{Gartner2023}. To ensure high data quality, data cleaning is an important and necessary step. At the same time, data cleaning still requires a great deal of manual effort and is one of the most time-consuming tasks for data scientists~\cite{Mahdavi2020}. For this reason, there has been extensive development in the fields of data cleaning algorithms and tools~\cite{Boehm2019}. Test data is required to evaluate new approaches and compare them with existing solutions. 

While real data is important for covering real scenarios, synthetic data is also highly significant for many reasons. These include privacy (especially in the health care domain), security and availability (some data cannot be accessed, or their retrieval is very time-consuming)~\cite{Murtaza2023, Endres2022}. Moreover, real data is often insufficient to cover all possible scenarios, such as certain distributions or error types. Synthetic data can be used to address specific test cases and approaches for many different domains. Thus, generating synthetic data is quite important for researchers and business developers~\cite{Endres2022}. In addition, common real-world datasets used in data quality research are already known to LLMs. Therefore, new data is necessary for a fair comparison of new methods. In this context, the importance of synthetic data is also increasing.~\cite{Abedjan2025}

As a contribution to this, we have developed GouDa, a tool for the automated generation of data sets with the possibility to include specific error types. We first introduced GouDa in a workshop paper with the title \textit{GouDa -- Generation of universal Data Sets: Improving Analysis and Evaluation of Data Preparation Pipelines}~\cite{GouDa}. This article is an extended version of this paper. The present version is more comprehensive with a stronger focus on data generation, errors, and data quality. In addition, we have further developed GouDa since the last publication to GouDa 2.0, making it more user-friendly and easier to create data sets.

\newpage

In summary, our contribution is as follows:
\begin{itemize}
    \item With GouDa 2.0, we provide a powerful data generator for synthetic data.
    \item A graphical user interface (GUI) simplifies data generation.
    \item Custom lists with possible attribute values enable the generation of data that is as realistic as possible.
    \item GouDa offers the option of inserting controlled errors into the data and generating both the erroneous and the clean data.
    \item With GouDa, both tabular and semi-structured data (for NoSQL data support) can be generated.
    \item GouDa enables the creation of multiple entity types and the realization of complex relationships between them.
    \item By using GouDa schema files, data can be generated in a reproducible manner.
\end{itemize}
GouDa thus offers the possibility to generate flexible and reproducible data sets with and without errors.  In this way, we aim to contribute to a common benchmark with standardized data sets for reproducible data quality research. In the following, we will go into more detail on the above points.

The paper is structured as follows: Section~\ref{sec:relatedwork} presents related work. Section~\ref{sec:GouDa} provides a detailed presentation of GouDa, including data generation, error types, architecture, and usage. In Section~\ref{sec:example}, we use an example to show how real data sets can be simulated. Section~\ref{sec:conclusion} summarizes the paper and provides an outlook on future work.

\section{Related Work}
\label{sec:relatedwork}
Many test data generators exist, for example~\cite{Galler2014, Ping2017, Castelein2018, Jatana2020, faker}, in both academic and commercial contexts. However, these generators focus on other areas, like software testing or SQL queries. The functionality to generate defined data errors with defined error rates is not provided.

Arocena et al.~\cite{Arocena2015} developed a tool for generating data sets with defined errors -- \textit{BART (Benchmarking Algorithms for data Repairing and Translation)}. With this tool, errors are inserted into clean databases that are already in place, thus focusing on constraint-induced errors. In addition, a limited selection of other data errors can be inserted (typos, duplicates, bogus or null values, and outliers). BART requires a relational database schema with unique tuple identifiers and therefore does not support other database types, such as NoSQL databases.

Another tool to obtain data as realistic as possible, is \textit{tab\_err}~\cite{Jung2025}. It can be used to insert errors into existing data. Various error types are available, such as missing values, outliers, typos, wrong units, and more. However, as with BART, tab\_err is limited to tabular data. An existing data set is always required.

One way to generate data for NoSQL databases is offered by \textit{Deimos}~\cite{Chillon2020}. Deimos is a model-based language developed to generate synthetic data from NoSQL schemas. It supports the variability of NoSQL data, such as structural variations, aggregations, and references with referential integrity. It also offers the option of inserting errors into the data. However, only a few types of error are supported, such as missing values, incorrect data types, and duplicates. Synthetic data can be generated, but it is also possible to reuse parts of existing data sets. Unfortunately, no publicly available code could be found.

With GouDa, we present the only data generator that
\begin{itemize}
    \item does not require existing data sets,
    \item still generates data as realistic as possible,
    \item supports a wide range of error types,
    \item and can generate both tabular data and data for NoSQL databases.
\end{itemize}

\section{GouDa}
\label{sec:GouDa}
Unlike BART~\cite{Arocena2015} and tab\_err~\cite{Jung2025}, GouDa does not need existing data. Instead, we create new JSON data sets and thus achieve a high flexibility. The basis for GouDa is the data generator implemented as part of the EvoBench project~\cite{Conrad2021}. This data generator is based on the publicly available library \textit{json-data-generator}\footnote{\url{https://github.com/vincentrussell/json-data-generator/tree/json-data-generator-1.12}}. This library can be extended by \textit{generators}. For example, functionality to define relations between different entity types was added as part of the EvoBench project~\cite{Conrad2021}. As part of the GouDa project, the ability to generate errors was added.

After the publication of the first version~\cite{GouDa}, we further developed GouDa. As part of this process, a graphical user interface (GUI) was implemented. This simplifies data generation. Also, an abstract data model was developed. It represents the JSON data transmitted by the GUI as Java objects. This allows for easier extensibility and validation. The generation of multi data sets (multiple entity types) has also been improved, so that more complex relationship types are now possible.

In the further course of the chapter, we'll first describe how GouDa can be used to generate data. Then we'll take a closer look at the types of errors it supports. After that, we'll introduce the architecture and go over the usage of GouDa.

\subsection{Data Generation}
The basis for data generation is provided by \textit{GouDa Schema Definitions (GSD)}. These are templates represented in JSON format. By using the JSON format, not only relational data but also nested data is supported. In a GSD, the structure of the JSON data to be generated and the respective values are defined. Moreover, relationships can be specified in different distributions and the number of objects to be created. Internal and external references allow values from other fields to be reused.

To generate data, values can be specified either statically or dynamically using GouDas generator functions. These can be divided into three categories: \textit{plain generators}, \textit{dictionary generators}, and \textit{error generators}. 

Plain generators convert input parameters into output parameters using Java functions. An example is the generator to create random doubles:
\begin{lstlisting}[language=Python]
    double(1.00, 99.99, "%.2f")
\end{lstlisting}

Here, random doubles are generated, in the range from $[1.00, 99.99)$, with two decimal places.

Dictionary generators choose random values from a dictionary, which is usually provided as a text file:
\begin{lstlisting}[language=Python]
    brands()
\end{lstlisting}

This function generates brand names. The corresponding names are taken from a text file. If necessary, custom lists and generators can be added here. This makes GouDa easily extendable and flexible in use.

Error generators can be used to insert controlled errors. The corresponding error types (see Section~\ref{subsec:error-types}) and rates are configurable:
\begin{lstlisting}[language=Python]
    error("MISSINGVALUE", 10, brands())
\end{lstlisting}

Here, the function from the previous example -- \texttt{brands()} -- is used to create brand names. Then, 10\% of the names are removed, generating a data set with errors. In addition, a clean data set is generated for each data set with errors (for example including the missing brand names).

The GUI simplifies data creation because all available generators are displayed in a drop-down menu. You can assign a name to a field, select a generator, and optionally add errors. If multiple entity types are created, there is an extra view for relationships. Here you can select the type of relationship to be used. The following relationship types are possible:
\begin{itemize}
    \item One-To-One
    \item One-To-Many
    \item Many-To-Many
\end{itemize}
Furthermore, the distribution can be determined. The following options are available for this purpose:
\begin{itemize}
    \item Uniform
    \item PoissonMax
    \item GaussianMinMax
    \item BetaMax
    \item Binomial
\end{itemize}

By default, the data is generated in the main memory and then stored on the hard disk. However, it is also possible to generate the data directly in a MongoDB database. Support for other databases will follow in future versions.

In summary, data generation with GouDa offers the following key benefits:
\begin{itemize}
    \item \textbf{Realistic}: For better applicability, synthetic data sets should be as realistic as possible. To achieve this, the opportunity to include lists with possible attribute values is provided. Different lists with realistic names, words or addresses are already included. Your own new lists can also be added. Our goal is to obtain data sets that are as close as possible to real data sets while still benefiting from the advantages of synthetic data generation.
    \item \textbf{Portable}: As mentioned, the data sets are generated in JSON format. This can be used for different systems. This has the advantage that not only tabular data can be generated, but also data for NoSQL databases.
    \item \textbf{Reproducible}: By sharing the schema definition, others can create the same data. 
    \item \textbf{Scalable}: Various options for scaling are available. This means that even large amounts of data can be used. This makes it possible to examine the scalability of data cleaning tools. Data quality is thus ensured even in applications with large amounts of data.
    \item \textbf{Error types}: A variety of different error types can be generated. These are shown in Table~\ref{tab:errortypes} and described in greater detail in the next section. Due to the comprehensive error definition, the data can be tailored precisely to the use case. It allows for many different aspects of data quality to be taken into account.
    \item \textbf{Error rate}: The occurrence rate is freely configurable for each error, the errors are distributed randomly. It is also planned to include other error mechanisms~\cite{Little2019} such as \textit{Missing At Random (MAR)} and \textit{Missing Not At Random (MNAR)}\/ in a future version. The freely adjustable error rate allows the impact of errors on subsequent analyses to be examined more closely.
    \item \textbf{Ground truth}: Ground truth is provided. The generated errors are additionally logged in a file. This allows for a more detailed examination of data quality. It is possible to determine how close the cleaning is to the ground truth. It is also possible to analyze whether all errors could be corrected. This represents a major advantage over real data, where ground truth is usually unavailable or can only be created with great effort.
\end{itemize}

\begin{table}[h]
  \centering
  \caption{Error types supported by GouDa}
  \label{tab:errortypes}
  \begin{tabular}{ll}
    \toprule
    Level & Error Type\\
    \midrule
    \multirow{9}*{\shortstack[l]{An Attribute Value\\ of a Single Tuple}} & Missing value\\
    \cline{2-2}
     & Syntax violation\\
    \cline{2-2 }
     & Interval violation\\
    \cline{2-2 }
     & Set violation\\
    \cline{2-2 }
     & Misspelled error\\
    \cline{2-2 }
     & Inadequate value to the attribute context\\
    \cline{2-2 }
     & Value items beyond the attribute context\\
    \cline{2-2 }
     & Meaningless Value\\
     \cline{2-2 }
     & Erroneous entry*\\
    \hline
    \multirow{4}*{\shortstack[l]{The Values\\ of a Single Attribute}} & Uniqueness value violation\\
    \cline{2-2}
     & Synonyms existence\\
    \cline{2-2 }
     & Outlier*\\
     \cline{2-2 }
     & Missing Attribute*\\
    \hline
    \multirow{3}*{\shortstack[l]{The Attribute Values\\ of a Single Tuple}} & Semi-empty tuple\\
    \cline{2-2}
     & Inconsistency among attribute values\\
    \cline{2-2 }
     & Irrelevant observation*\\
    \hline
    \multirow{4}*{\shortstack[l]{The Attribute Values\\ of Several Tuples}} & Redundancy about an entity\\
    \cline{2-2}
     & Inconsistency about an entity\\
    \cline{2-2}
     & Bias*\\
    \cline{2-2 }
     & Noise*\\
  \bottomrule
\end{tabular}
\end{table}

\subsection{Error Types}
\label{subsec:error-types}
As described, errors can be added to the data in a controlled manner using error generators. Many different definitions of error types are presented in literature~\cite{Rahm2000, Kim2003, Mueller2005, Oliveira2005, Li2011}. The classification of GouDa (see \autoref{tab:errortypes}) is based on the definition of Oliveira et al.~\cite{Oliveira2005}. This definition follows a bottom-up approach in which the errors are classified into six different levels. 
These levels can be divided into several groups: A distinction is made between \textit{Single Data Source} and \textit{Multiple Data Sources} (comparable to the definition of~\cite{Rahm2000}). The group of  \textit{Single Data Source} can be further divided into \textit{Single Relation} and \textit{Relationships among Multiple Relations}. The group \textit{Single Relation} includes the four lowest levels \textit{An Attribute Value of a Single Tuple}, \textit{The Values of a Single Attribute}, \textit{The Attribute Values of a Single Tuple}, and \textit{The Attribute Values of Several Tuples.} The approach is well suited to the controlled generation of errors because of the bottom-up approach. Error generation was first started on the lowest level (\textit{An Attribute Value of a Single Tuple}). Then other levels were added step by step.

The definition by Oliveira et al.~\cite{Oliveira2005} is already very comprehensive and includes a large part of error types from other works. In addition, the classification has been extended to include further important error types, based on~\cite{Rahm2000, Kim2003, Mueller2005, Li2011}. This concerns e.g.\ outlier and bias. The corresponding types are marked with a * in \autoref{tab:errortypes}. Some definitions from \cite{Oliveira2005} were not taken into account. One example is \textit{Outdated Value}. The goal of GouDa is to create a static data set. In this context, an outdated value would not be recognizable. For the same reason, the last level (\textit{Multiple Data Sources}) of \cite{Oliveira2005} is not considered. The focus is only on static data sets and not on errors caused by merging multiple data sources. This is intended for a future version of GouDa.

To ensure that the error types reflect real-world conditions, we analyzed real-world data from a global pharmaceutical and diagnostics company. Table~\ref{tab:errortypes-company} shows the occurring error types and their frequency. The analysis shows that the errors we selected for GouDa are not only cited in various scientific papers~\cite{Rahm2000, Kim2003, Mueller2005, Oliveira2005, Li2011} but also occur in practice.

\begin{table}[ht]
  \centering
  \caption{Occurrence and frequency of data errors in applications of the information system at a production site of a global pharmaceutical and diagnostics company. No frequency was specified for the error type \textit{synonyms existence}, as there was no official entry indicating which term is the correct one.}
  \label{tab:errortypes-company}
  \begin{tabular}{lllr}
    \toprule
    System & Data Storage & Error Type & Frequency\\
    \midrule
    System A & MS SQL Server & Missing value & 6.9\%\\
     &  & Redundancy about an entity & 5.7\%\\
     \midrule
    System B & Asset Framework DB & Missing value & 0.1\%\\
    &  & Erroneous entry & 11.5\%\\
    \midrule
    System C & MS SQL Server & Erroneous entry & 0.1\%\\
     &  & Inconsistency about an entity & 0.2\%\\
     \midrule
    Multiple & Multiple & Synonyms existence &  \\
  \bottomrule
\end{tabular}
\end{table}

\subsection{Architecture and Usage}
\begin{figure}[ht]
  \centering
  \includegraphics[width=0.75\linewidth]{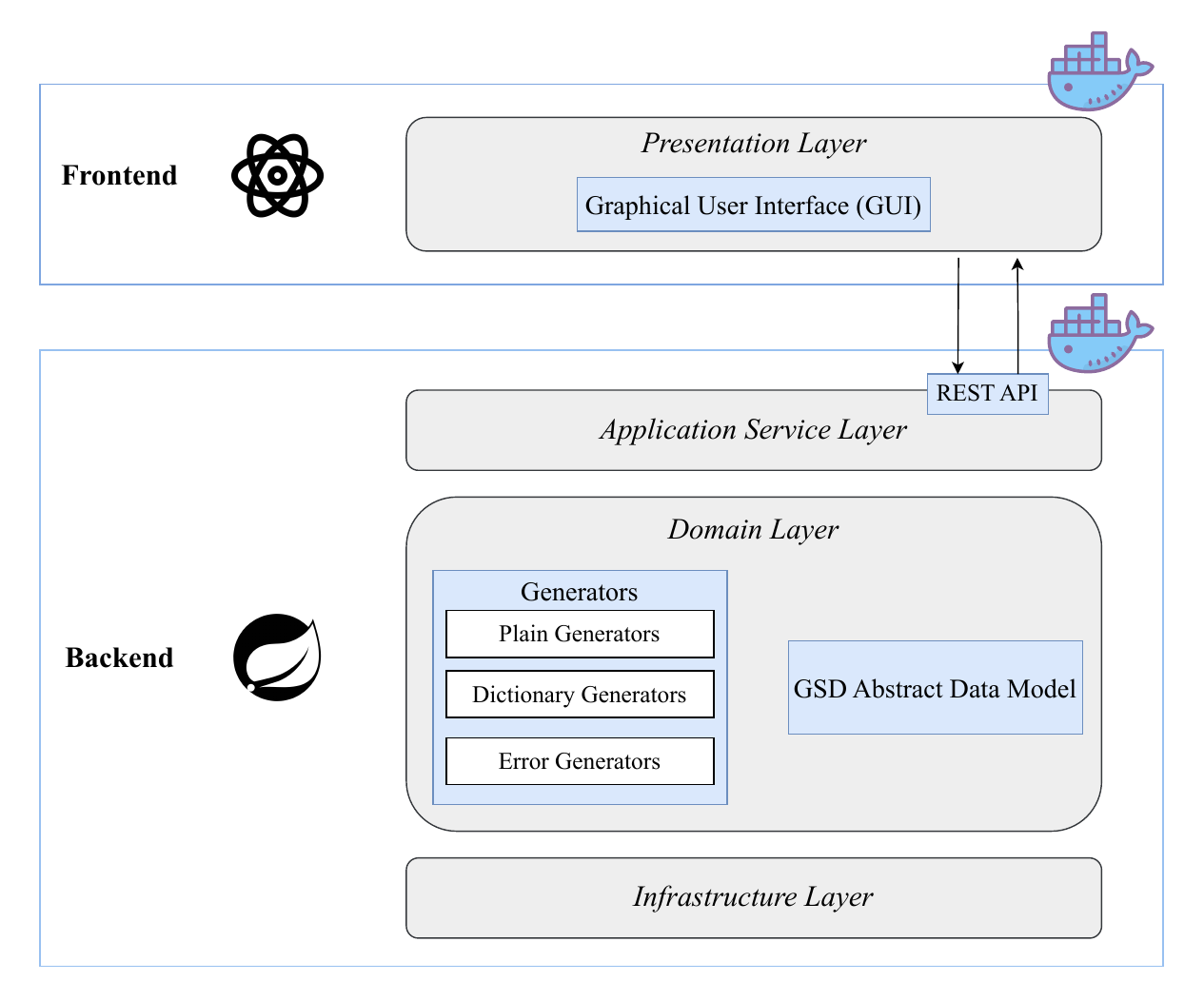}
  \caption{GouDa High Level Architecture}
  \label{fig:gouda-architecture}
\end{figure}

GouDa implements a monolithic, technically partitioned layered architecture. Its four main layers are the \textit{presentation layer}, the \textit{application service layer}, the \textit{domain layer}, and the \textit{infrastructure layer}. Additionally a layer for \textit{cross-cutting concerns} provides utilities such as logging and performance monitoring. An overview of the high level architecture is shown in \autoref{fig:gouda-architecture}. 

As can be seen, GouDa is provided with two Docker containers, one for the frontend and one for the backend. The frontend contains the presentation layer, where the GUI is located. The implementation is done with React\footnote{\url{https://react.dev/}}. The frontend communicates with the backend via a REST API, located in the application service layer. The domain layer and infrastructure layer are also part of the backend. The domain layer contains the generators described above and the abstract data model, which is used for validation. The backend is realized with Spring Boot\footnote{\url{https://spring.io/projects/spring-boot}}. 

Therefore the only dependencies for using GouDa are \textit{Docker} and \textit{Docker Compose}. To get GouDa up and running first the GouDa Git repository\footnote{\url{https://gitlab.com/cheese-board/gouda}} needs to be cloned. After that, within the main directory, the following command can be executed to start the application: 
\begin{lstlisting}
    docker compose up
\end{lstlisting}
This will start both the frontend and the backend. The GUI can then be accessed via \url{http://localhost}. 

\section{Example of GouDa in use}
\label{sec:example}
In this section, we use an example to show how data can be generated with GouDa.  We will show how the generators available in GouDa can be used to flexibly generate synthetic data.
A complete list of all generators (including error generators) can be found here: \url{https://gitlab.com/cheese-board/gouda}. 
 
GouDa can be used to generate synthetic data in a variety of ways. In particular, realistic data can also be recreated. This approach has various advantages, for example, in terms of scaling and the controlled insertion of error types and error rates. In this chapter, we illustrate these possibilities using an example and discuss the specific possibilities and advantages of this approach. 

To highlight the advantages of synthetic data while remaining as realistic as possible, we have recreated a real-world data set about beers. This data set is used by Mahdavi et al.~\cite{raha}, among others, in the context of error detection research. They scraped the data from the web, cleaned it manually, and then inserted errors. Both the erroneous (\textit{dirty.csv}) and the clean (\textit{clean.csv}) versions of the data are publicly available~\footnote{\url{https://github.com/BigDaMa/raha/tree/master/datasets/beers}}.
The data set contains the following attributes:
\begin{itemize}
    \item \texttt{id}: A random ID (integer), identifying the record (integer) -- for this purpose, we simply chose a sequential number. This is included by default in every data record generated with GouDa.
    \item \texttt{beer-name}: The name of the beer (text) -- we extracted all distinct beer names from the \textit{clean.csv} file and added them as a custom list. Even though we have used real data as a reference here, no real data is required for the use of GouDa as described. The lists could thus also be generated manually, by an LLM, or by other means. As mentioned, GouDa already comes with various lists (e.g., for names or brands).
    \item \texttt{style}: The style of the beer (text) -- we extracted all styles from the \textit{clean.csv} file and add a custom map with the \texttt{beer-name} as key and the \texttt{style} as value. This way, we can add the correct styles to the beer names. Here, again, the use of an LLM would be a viable alternative. 
    \item \texttt{ounces}: Size of the beer (float) -- for this purpose, we selected the distinct values of the attribute from \textit{clean.csv}. Since there was only a set of seven distinct values, we chose a generator that selects randomly from this set. Here, it would also be possible to make the data generation even more realistic and generate values depending on the \texttt{style} for example.
    \item \texttt{abv}: Alcoholic content by volume (float) -- for this purpose, we chose the function that randomly generates floats. We extracted the min and max values of the attribute from \textit{clean.csv} and passed them to the generator. Here again, it would also be possible to make the data generation even more realistic and generate values depending on the \texttt{style} for example.
    \item \texttt{ibu}: International bitterness unit (float) -- here we have chosen the same approach as for \texttt{abv}.
    \item \texttt{brewery}: The name of the brewery (text) -- comparable to \texttt{beer-name}, we extracted all distinct brewery names from \textit{clean.csv} and added them as a custom list.
    \item \texttt{brewery-id}: The ID of the brewery (integer) -- comparable to \texttt{style}, we created a custom map with \texttt{brewery} as key and \texttt{brewery-id} as value. Alternatively, a simpler approach would also be conceivable, whereby each brewery is simply given a random unique id.
    \item \texttt{city}: The city, where the brewery is located (text) -- in order to obtain the correct city names, we created a custom map with the \texttt{brewery} as key and the \texttt{city} as value.
    \item \texttt{state}: The corresponding state (text) -- again, we used a custom map with the \texttt{city} as key and the \texttt{state} as value, to obtain the correct states to the city.
\end{itemize}

The paper by Mahdavi et al.~\cite{raha} states that the erroneous data set contains missing values, formatting issues, and violated attribute dependency. Applied to the error classification of GouDa, we have therefore included the following errors:
\begin{itemize}
    \item \textit{Missing values} in 5\% of \texttt{beer-name}
    \item \textit{Missing values} in 10\% of \texttt{ibu}
    \item \textit{Wrong units} in 15\% of \texttt{ounces} -- the correct version is the value itself, e.g.\ \texttt{12.0} and for the errors we added the unit: \texttt{12.0 oz}
    \item \textit{Wrong units} in 15\% of \texttt{abv} -- the correct version is the value itself, e.g.\ \texttt{0.005} and for the errors we added the unit: \texttt{0.005\%}
    \item \textit{Derived value error} in 5\% of \texttt{style}: The style is dependent of the \texttt{beer-name}, for 5\% we added the wrong \texttt{style}
    \item \textit{Derived value error} in 5\% of \texttt{state}: The state is dependent of the \texttt{city}, for 5\% we added the wrong \texttt{state}
\end{itemize}

The GouDa Schema Definition (GSD) looks as follows:
\begin{lstlisting}[language=Python]
{
  "index": "{{Index(Index_1(0))}}",
  "beer-name": "{{error(\"MISSINGVALUE\", 5,put(\"beer-name\",customSet(\"beer-names.txt\")))}}",
  "style": "{{error(\"DERIVED_VALUE_ERROR\", 5, \"beer-styles.txt\",\"beer-name\")}}",
  "ounces": "{{error(\"WRONG_UNIT\", 15, random(12.0, 8.4, 16.0, 24.0, 19.2, 32.0, 16.9), \" oz\")}}",
  "abv": "{{error(\"WRONG_UNIT\", 15, double(0.001,0.1,\"%.3f\"), \"%\")}}",
  "ibu": "{{error(\"MISSINGVALUE\", 10, double(4.0,138.0,\"%.0f\"))}}",
  "brewery": "{{put(\"brewery\", customSet(\"breweries.txt\"))}}",
  "brewery-id": "{{customDerivedValue(\"brewery-map.txt\",get(\"brewery\"))}}",
  "city": "{{put(\"city\", customDerivedValue(\"brewery-city.txt\",get(\"brewery\")))}}",
  "state": "{{error(\"DERIVED_VALUE_ERROR\", 5, \"city-state.txt\",\"city\")}}",
  "Relationship": "{{Relationship(None,Relationship_1,Index_1,100)}}"
}
\end{lstlisting}

By using the GUI, this can be generated without requiring a deeper understanding of the syntax of the GSD. In the following, we show an exemplary excerpt of the data

\begin{minipage}{0.48\textwidth}
\begin{lstlisting}[language=Python]
// erroneous version
{
    "index": "33",
    "beer-name": "",
    "style": "American Amber / Red Ale",
    "ounces": "16.9 oz",
    "abv": "0.038%",
    "ibu": 133,
    "brewery": "Bold City Brewery",
    "brewery-id": 378,
    "city": "Jacksonville",
    "state": "PA"
}
\end{lstlisting}
\end{minipage}
\hfill
\begin{minipage}{0.48\textwidth}
\begin{lstlisting}[language=Python]
// ground truth
{
    "index": "33",
    "beer-name": "Bomber Mountain Amber Ale (2013)",
    "style": "American Amber / Red Ale",
    "ounces": 16.9,
    "abv": 0.038,
    "ibu": 133,
    "brewery": "Bold City Brewery",
    "brewery-id": 378,
    "city": "Jacksonville",
    "state": "FL"
}
\end{lstlisting}
\end{minipage}

The incorrect version is shown on the left. There are \textit{missing values} in \texttt{beer-name}, \textit{formatting issues} (wrong unit) in \texttt{ounces} and \texttt{abv}, and a \textit{violated attribute dependency} between \texttt{city} and \texttt{state}. The right side shows the ground truth with the correct values.

\paragraph{Discussion} As can be seen, with GouDa we were able to recreate the data set very realistically. Even though the data was generated as JSON, a flat structure was created so that the data can be read in with \texttt{pd.read\_json()} just as it would be with the pandas function \texttt{pd.read\_csv()}. This emphasizes the flexibility of GouDa. Analysis of the \textit{clean.csv} data set revealed the advantages of synthetically generated data. The data set was originally cleaned manually, but the following errors were still identified in the data:
\begin{itemize}
    \item There is an entry for the city \textit{Marquette} with the state \textit{MA}. This is incorrect; the correct state is \textit{MI}.
    \item The Brewery with the name \textit{Against the Grain Brewery} appears in different spellings, twice as \textit{Against The Grain Brewery} (with capital T) and 13 times as \textit{Against the Grain Brewery} (with lowercase t). All entries refer to the city Louisville, but the two breweries have different values for \texttt{brewery-id}. The correct value is \textit{Against the Grain Brewery} (with lowercase t).
\end{itemize}
This shows that manual cleaning of data records is prone to errors. With GouDa, realistic data can be generated in a controlled manner so that only the intended errors are present in the data. The generation of synthetic data also offers further advantages: Depending on requirements, the existing error rates can easily be adjusted and additional errors can be added. The size of the data can also be adjusted as desired, so that the data can be easily used for scalability tests. If additional data types (e.g., date values) are required, these can also be added as additional attributes without much effort. It is also possible to split the data set into several data sets, e.g.\ by storing the breweries as a separate entity type. The appropriate relationship type can then be selected from the available options (e.g., N:M). In addition to recreating real data, purely synthetic data can also be created using GouDas numerous generators. Real data as a basis is not necessary. In summary, this example shows how GouDa can be used to leverage the advantages of synthetic data generation while still ensuring that the data remains realistic.

\section{Conclusion and Future Work}
\label{sec:conclusion}
In this paper, we have presented our data generator GouDa. GouDa is flexible in its application: the use of JSON enables support for many different data formats. A wide range of generator functions allows many different types of data to be created. In addition, you can add your own lists of possible attribute values and key-value pairs. This allows data to be generated in a realistic manner. Furthermore, it is possible to add controlled errors to the data while simultaneously generating an error-free version of the data. Multiple entity types can be generated, which can be linked to each other using different relationship types.

All these features make GouDa a powerful data generator capable of generating FAIR data. This is particularly important for data quality research~\cite{Jung2025}. Take, for example, the generated synthetic \textit{beers} dataset: The data is F(indable) and A(ccessible)\footnote{\url{https://gitlab.com/cheese-board/gouda}}. By using GSDs, which are JSON files, the data set is also I(nteroperable). The GSD is R(eusable) and can thus also be used by others to generate the data.

Using an example, we have highlighted the advantages of generating synthetic data and demonstrated how GouDa can be used flexibly. In this way, we aim to contribute to a common benchmark of standardized data sets for reproducible data quality research.

We are continuously developing GouDa so that other error mechanisms (such as MAR) will soon be possible. We also plan to add more generators to enable the creation of even more diverse data. As described, support for additional databases besides MongoDB is also expected to be added in the near future. Furthermore, we plan to add data generation for multiple data sources. 





\section*{Acknowledgment}
GouDa would not be what it is today if we hadn't worked with such great students: Thanks to Gerrit Boerner for most of the implementation work related to error generation. Thanks to the group of the \textit{Fachpraktikum} from the winter semester 23/24, who created the initial GUI and the abstract data model, among other things. Many thanks to Martin Schmid, who merged the different versions, added tests, and created a comprehensive concept on the topic of software development in research contexts.

\bibliographystyle{alpha}
\bibliography{references}

\newcommand{\etalchar}[1]{$^{#1}$}
\begin{thebibliography}{AGM{\etalchar{+}}15}

\bibitem[Abe25]{Abedjan2025}
Ziawasch Abedjan.
\newblock Navigating disruption: The impact of {AI} technologies on data integration research.
\newblock Keynote at the 2nd International Workshop on Data-Centric {AI} (DATAI), co-located with VLDB 2025, 2025.

\bibitem[AGM{\etalchar{+}}15]{Arocena2015}
Patricia~C. Arocena, Boris Glavic, Giansalvatore Mecca, Ren{\'{e}}e~J. Miller, Paolo Papotti, and Donatello Santoro.
\newblock Messing up with {BART:} error generation for evaluating data-cleaning algorithms.
\newblock {\em Proc. {VLDB} Endow.}, 9(2):36--47, 2015.

\bibitem[BKY19]{Boehm2019}
Matthias Boehm, Arun Kumar, and Jun Yang.
\newblock {\em Data Management in Machine Learning Systems}.
\newblock Synthesis Lectures on Data Management. Morgan {\&} Claypool Publishers, 2019.

\bibitem[CAS{\etalchar{+}}18]{Castelein2018}
Jeroen Castelein, Maur{\'{\i}}cio~Finavaro Aniche, Mozhan Soltani, Annibale Panichella, and Arie van Deursen.
\newblock Search-based test data generation for {SQL} queries.
\newblock In {\em {ICSE}}, pages 1220--1230, New York, NY, 2018. {ACM}.

\bibitem[CMK{\etalchar{+}}21]{Conrad2021}
Andr{\'{e}} Conrad, Mark~Lukas M{\"{o}}ller, Tobias Kreiter, Jan{-}Christopher Mair, Meike Klettke, and Uta St{\"{o}}rl.
\newblock Evobench: Benchmarking schema evolution in nosql.
\newblock In {\em {TPCTC}}, volume 13169 of {\em Lecture Notes in Computer Science}, pages 33--49, Heidelberg, 2021. Springer.

\bibitem[CRM20]{Chillon2020}
Alberto~Hern{\'{a}}ndez Chill{\'{o}}n, Diego~Sevilla Ruiz, and Jes{\'{u}}s~Garc{\'{\i}}a Molina.
\newblock Deimos: {A} model-based nosql data generation language.
\newblock In {\em {ER} (Workshops)}, volume 12584 of {\em Lecture Notes in Computer Science}, pages 151--161, Heidelberg, 2020. Springer.

\bibitem[EN25]{Ehrlinger2025}
Lisa Ehrlinger and Felix Naumann.
\newblock {\em Data Quality for Enterprise AI}, pages 91--128.
\newblock Springer Nature Switzerland, Cham, 2025.

\bibitem[EVT22]{Endres2022}
Markus Endres, Asha~Mannarapotta Venugopal, and Tung~Son Tran.
\newblock Synthetic data generation: {A} comparative study.
\newblock In {\em {IDEAS}}, pages 94--102, New York, NY, 2022. {ACM}.

\bibitem[GA14]{Galler2014}
Stefan~J. Galler and Bernhard~K. Aichernig.
\newblock Survey on test data generation tools - an evaluation of white- and gray-box testing tools for c{\#}, c++, eiffel, and java.
\newblock {\em Int. J. Softw. Tools Technol. Transf.}, 16(6):727--751, 2014.

\bibitem[jF26]{faker}
{joke2k} and {Faker contributors}.
\newblock Faker: Python package that generates fake data for you, 2026.
\newblock Python package, accessed 22 July 2026.

\bibitem[JJCB25]{Jung2025}
Philipp Jung, Sebastian J{\"a}ger, Nicholas Chandler, and Felix Biessmann.
\newblock Towards realistic error models for tabular data.
\newblock {\em ACM Journal of Data and Information Quality}, 17(4):1--27, 2025.

\bibitem[JS20]{Jatana2020}
Nishtha Jatana and Bharti Suri.
\newblock An improved crow search algorithm for test data generation using search-based mutation testing.
\newblock {\em Neural Process. Lett.}, 52(1):767--784, 2020.

\bibitem[KCH{\etalchar{+}}03]{Kim2003}
Won~Y. Kim, Byoung{-}Ju Choi, Eui~Kyeong Hong, Soo{-}Kyung Kim, and Doheon Lee.
\newblock A taxonomy of dirty data.
\newblock {\em Data Min. Knowl. Discov.}, 7(1):81--99, 2003.

\bibitem[LPK11]{Li2011}
Lin Li, Taoxin Peng, and Jessie Kennedy.
\newblock A rule based taxonomy of dirty data.
\newblock {\em GSTF Journal on Computing (JoC)}, 1(2), 2011.

\bibitem[LR19]{Little2019}
Roderick~JA Little and Donald~B Rubin.
\newblock {\em {Statistical analysis with missing data}}, volume 793.
\newblock John Wiley \& Sons, 2019.

\bibitem[MA20]{Mahdavi2020}
Mohammad Mahdavi and Ziawasch Abedjan.
\newblock Baran: Effective error correction via a unified context representation and transfer learning.
\newblock {\em Proc. {VLDB} Endow.}, 13(11):1948--1961, 2020.

\bibitem[MAF{\etalchar{+}}19]{raha}
Mohammad Mahdavi, Ziawasch Abedjan, Raul~Castro Fernandez, Samuel Madden, Mourad Ouzzani, Michael Stonebraker, and Nan Tang.
\newblock Raha: {A} configuration-free error detection system.
\newblock In {\em {SIGMOD} Conference}, pages 865--882, New York, NY, 2019. {ACM}.

\bibitem[MAK{\etalchar{+}}23]{Murtaza2023}
Hajra Murtaza, Musharif Ahmed, Naurin~Farooq Khan, Ghulam Murtaza, Saad Zafar, and Ambreen Bano.
\newblock Synthetic data generation: State of the art in health care domain.
\newblock {\em Comput. Sci. Rev.}, 48:100546, 2023.

\bibitem[MF03]{Mueller2005}
Heiko M{\"u}ller and Johann~Christoph Freytag.
\newblock Problems, methods, and challenges in comprehensive data cleansing.
\newblock Technical report hub-ib-164, Humboldt University, 2003.

\bibitem[ORHG05]{Oliveira2005}
Paulo Oliveira, F{\'a}tima Rodrigues, Pedro Henriques, and Helena Galhardas.
\newblock A taxonomy of data quality problems.
\newblock In {\em 2nd Int. Workshop on Data and Information Quality}, pages 219--233, Heidelberg, 2005.

\bibitem[PSH17]{Ping2017}
Haoyue Ping, Julia Stoyanovich, and Bill Howe.
\newblock Datasynthesizer: Privacy-preserving synthetic datasets.
\newblock In {\em {SSDBM}}, pages 42:1--42:5. {ACM}, 2017.

\bibitem[RBCS22]{GouDa}
Valerie Restat, Gerrit Boerner, Andr{\'{e}} Conrad, and Uta St{\"{o}}rl.
\newblock Gouda - generation of universal data sets: Improving analysis and evaluation of data preparation pipelines.
\newblock In {\em DEEM@SIGMOD}, pages 2:1--2:6, New York, NY, 2022. {ACM}.

\bibitem[RD00]{Rahm2000}
Erhard Rahm and Hong~Hai Do.
\newblock Data cleaning: Problems and current approaches.
\newblock {\em {IEEE} Data Eng. Bull.}, 23(4):3--13, 2000.

\bibitem[Res23]{Gartner2023}
Gartner Research.
\newblock How poor data quality costs organizations \$12.9 million annually.
\newblock Technical report, Gartner, 2023.

\end{thebibliography}

\end{document}